\documentclass[twocolumn,preprintnumbers,amsmath,amssymb]{revtex4}

\usepackage{graphicx}
\usepackage{dcolumn}
\usepackage{bm}

\usepackage[usenames]{color}

\newcommand\be{\begin{equation}}
\newcommand\ba{\begin{eqnarray}}
\newcommand\ee{\end{equation}}
\newcommand\ea{\end{eqnarray}}

\newcommand\bw{\begin{widetext}}
\newcommand\ew{\end{widetext}}

\newcommand\nn{\nonumber}
\newcommand\mrm{\mathrm}

\newcommand\zdrift{\Delta_t z}
\newcommand\mcz{\mathcal{M}_z}

\newcommand\psiaccel{\Psi_{\mrm{acc}}(f)}

\newcommand\cum {X_H^{\mathrm{(cum)}}}

\newcommand\etal{\textit{et al.}}

\begin{document}


\title{Direct Measurement of the Positive Acceleration of the Universe and Testing
Inhomogeneous Models under Gravitational Wave Cosmology
}

\author{Kent Yagi}
\affiliation{%
Department of Physics, Kyoto University,
   Kyoto, 606--8502, Japan
}%

\author{Atsushi Nishizawa and Chul-Moon Yoo}
\affiliation{
Yukawa Institute for Theoretical Physics,
  Kyoto University,
  Kyoto 606--8502, Japan
}%



\begin{abstract}
One possibility for explaining the apparent accelerating expansion of 
the universe is that we live in 
the center of a spherically inhomogeneous universe. 
Although current observations cannot fully distinguish $\Lambda$CDM and 
these inhomogeneous models, 
direct measurement of the acceleration of the universe can be 
a powerful tool in probing them. 
We have shown that, 
if $\Lambda$CDM is the correct model, 
DECIGO/BBO would be able to detect 
the positive redshift drift 
(which is the time evolution of the source redshift $z$) 
in 3--5 year gravitational wave (GW) observations from neutron-star binaries, 
which enables us to rule out any Lema\^itre-Tolman-Bondi (LTB) void model 
with monotonically increasing density profile.
We may even be able to rule out any LTB model 
unless we allow unrealistically steep density profile at $z\sim 0$. 
This test can be performed with GW observations alone, without any reference to electromagnetic observations, and 
is more powerful than the redshift drift measurement using Lyman $\alpha$ forest.
\end{abstract}

\maketitle

\textit{Introduction}: 
If we assume that our universe is homogeneous and isotropic, 
current cosmological observations (e.g. type Ia supernovae (SNe)~\cite{riess}) 
indicate the accelerating expansion of the universe. 
Dark energy 
and modification of the gravitational theory are two candidates 
that can explain these observational results.
However, once we allow for 
the possibility that our universe has cosmological-scale spherical inhomogeneity, with the observer at the center, 
the observations can be explained 
without introducing the unknown dark energy or alternative theories of gravity.
In such models,  
the Copernican Principle is apparently violated and 
the expansion of the universe is not necessarily accelerating.
As such, the direct detection of the cosmic acceleration 
is very useful in testing them. 
Although there are several proposals (e.g. Refs.~\cite{loeb, setoDECIGO}) for the direct detection of the cosmic acceleration,  
it has not been observed yet.
%
The direct detection of the acceleration of the universe 
provides not only a key to solving the dark energy problem, 
which is one of the biggest challenges in cosmology, 
but also a critical test of the Copernican Principle, 
which is one of the most essential ``principles" in cosmology.

The simplest example of the inhomogeneous model is 
the Lema\^itre-Tolman-Bondi (LTB) spacetime, 
which is a spherically symmetric, 
dust
solution of the Einstein Equations.
If we live at the center of the LTB spacetime with a 
Gpc-scale void, 
the apparent acceleration of the universe
can be explained. 
The LTB model 
has been partially tested with cosmological observations such as, 
the cosmic microwave background, baryon acoustic oscillations, the kinetic Sunyaev-Zeldovich effect etc., but
it has not been completely ruled out yet 
(see e.g. Ref.~\cite{biswas}).
One of the fundamental difficulties 
in testing the LTB models is due to
the ambiguities in the primordial spectrum. 
In the standard cosmology, 
we mostly assume a scale invariant spectrum 
based on the inflationary paradigm that 
makes our universe homogeneous in the early epoch. 
However, once we introduce cosmological scale inhomogeneities 
and violate the Copernican Principle, 
there is no strong motivation 
to assume inflation in the early epoch. 
Therefore, observations which are not affected by 
the primordial spectrum are very crucial in testing a wide class of LTB models. 
One such observation is 
the redshift-distance relation of type-Ia SNe. 
%
However, it is known that one can construct a 
LTB void model 
that exactly reproduces the redshift-distance relation of the 
$\Lambda$CDM model~\cite{yoo:inverse}. 
Therefore, we need other observations that do not depend on the primordial information.

Redshift-drift measurement is the one that meets our 
demands~\cite{uzan:zdrift} (see also Refs.~\cite{yoo:inverse, quartin, yoo:zdrift}).
Redshift drift is the time evolution of the redshift due to 
the acceleration of the universe, hence its detection
means
the direct measurement of the 
cosmic acceleration. 

In the Friedmann-Lema\^itre-Robertson-Walker (FLRW) spacetime, 
the redshift drift is given as 
$\zdrift = H_0 \Delta t_o \left( 1+z-\frac{H(z)}{H_0} \right)$, 
where $\Delta t_o$ denotes the observation period, 
and $H_0$ and $H(z)$ are the Hubble parameter at present 
and at redshift $z$~\cite{loeb}, respectively. 
%
%
In the $\Lambda$CDM universe, 
$\zdrift$ is \textit{positive} in the range $z=0$--2~\cite{quartin}.
On the other hand, typical LTB models that can 
explain observations usually have \textit{negative} $\zdrift$ at any 
$z$ (see e.g. Ref.~\cite{quartin}).
Recently, Yoo \etal~\cite{yoo:zdrift} showed that if the matter density 
monotonically increases with the radial distance $r$, 
$\zdrift$ must be \textit{negative} for any $z>0$.
Furthermore, they have shown that for any LTB density profile, $d\zdrift/dz  < 0$
%
%
for $z \ll 1$.
This indicates that, unlike the redshift-distance relation, it is impossible to exactly reproduce 
$\zdrift$ of the $\Lambda$CDM model with LTB models.
Therefore, it is crucial to measure the sign of $\zdrift$ 
at $z<2$ (especially $z \ll 1$) in order to distinguish 
the $\Lambda$CDM and LTB models. 

The order of magnitude of $\zdrift$ is roughly given as the cosmic age 
divided by the observation time, hence 
$\zdrift \sim 10^{-10} \times {\rm (obs. ~time)}/{\rm 1yr}$. 
It is this tiny value that makes it difficult to measure $\zdrift$ with current technology. 
Recently, 
Quartin and Amendola~\cite{quartin} have shown that 
by measuring the shift 
of the Lyman $\alpha$ forest of 
quasar spectrum at $z=2$--5 with the proposed E-ELT instrument 
CODEX~\cite{liske} for 10 yrs, it will be possible to distinguish 
$\Lambda$CDM and typical LTB void models. 
%
%
%
%
However, 
CODEX would not be able to measure $\zdrift$ at low $z$ 
since the Lyman $\alpha$ forest can be measured from ground only 
at $z \geq 1.7$~\cite{liske}.
They can only test typical LTB models and not generic ones.

In this letter, we estimate how accurately we can measure 
the redshift drift with future gravitational wave (GW) interferometers.
It seems that DECIGO~\cite{setoDECIGO,kawamura2011} and BBO~\cite{bbo} 
are the only proposed detectors that can measure $\zdrift$ at $z \leq 2$. 
These are complementary observations 
to electromagnetic ones. 
We consider neutron-star (NS) binaries as GW sources, which are often called standard sirens and can be unique tools to probe the cosmic expansion~\cite{schutz,nishizawa,DECIGOcosmology}.
When the expansion is accelerating, 
we may find an additional phase shift in gravitational waveforms associated with
the redshift drift~\cite{setoDECIGO}. 
We assume that $\Lambda$CDM is the correct model 
and estimate whether we can tell the positivity 
of the redshift drift at low $z$ with GW observations.

Throughout this letter, we use the unit $G=c=1$.

\textit{Measuring the redshift drift with GWs}:
Let us first derive the correction to the GW phase due to the accelerating expansion of the universe.
We here consider a binary consisting of two bodies 
with masses $m_1$ and $m_2$.
We define $h(\Delta t)$ as the observed GW where $\Delta t \equiv t_c - t$ denotes the time to coalescence measured in the observer frame with $t_c$ representing the coalescence time.
The Fourier component of this waveform is written as 
\ba
\tilde{h}(f) &=& \int^{\infty}_{-\infty} dt \ e^{2\pi i f t} h(\Delta t) \nn \\
&=& e^{2\pi i f t_c} \int^{\infty}_{-\infty} d\Delta t \ e^{-2\pi i f \Delta t} h(\Delta t)\,.
\label{tilde-h1}
\ea
Now,  $\Delta t$ can be related to $\Delta T \equiv (1+z_c) \Delta t_e$ as~\cite{setoDECIGO, takahashinakamura} 
\be
\Delta t = \Delta T + X(z_c) \Delta T^2\,,
\label{deltat}
\ee
where $z_c$ is the source redshift at coalescence, $\Delta t_e$ is the time to coalescence measured in the source frame and $X(z)$ is the acceleration parameter defined as $X(z) \equiv \frac{H_0}{2} \left(1-\frac{H(z)}{(1+z) H_0} \right)$.
Notice that $X(z)$ is related to the redshift drift $\zdrift$ as $\zdrift= 2(1+z)\Delta t_o X(z)$ in FLRW spacetime.
Then, $h(\Delta t)$ can be re-expressed as a function of $\Delta T (\Delta t)$ as $h(\Delta t) = H(\Delta T (\Delta t))$, where $H(\Delta T)$ corresponds to GWs without cosmic acceleration.
By substituting this equation and Eq.~\eqref{deltat} into Eq.~\eqref{tilde-h1}, we get
\ba
\tilde{h}(f) & = &  e^{2\pi i f t_c} \int^{\infty}_{-\infty} d\Delta T' \ e^{-2\pi i f \Delta T'} H(\Delta T') \nn \\
& & \times e^{-2\pi if X(z_c) \Delta T'^2}\,.
\ea
By using the stationary phase approximation~\cite{cutlerflanagan}, $e^{-2\pi if X(z_c) \Delta T'^2}$ in the integrand can be pulled out of the integral with $\Delta T'$ replaced by $\Delta T(f)  = 5(8\pi \mcz f)^{-8/3} \mcz$~\cite{cutlerflanagan}, where  $\mathcal{M}_z\equiv M (1+z_c) \eta^{3/5}$ denotes the redshifted chirp mass with $M \equiv (m_1+m_2) $ and  $\eta \equiv m_1m_2/M^2$ representing the total mass and the symmetric mass ratio, respectively.
Then, we obtain
\be
\tilde{h}(f) = e^{i\psiaccel} \tilde{h}(f) \big|_{\mathrm{no \ accel}}\,,
\ee
where 
\ba
\psiaccel &\equiv & -2\pi f X(z_c) \Delta T(f)^2 \nn \\
&=&  -\Psi_N (f) \frac{25}{768} X(z_c) \mcz x^{-4}\,,
\ea
with $x\equiv (\pi \mcz f)^{2/3}$ and $\Psi_N(f) \equiv \frac{3}{128}(\pi \mathcal{M}_z f)^{-5/3}$, and 
\be
\tilde{h}(f) \big|_{\mathrm{no \ accel}} =  e^{2\pi i f t_c} \int^{\infty}_{-\infty} d\Delta T' \ e^{-2\pi i f \Delta T'} H(\Delta T')\,,
\ee
which corresponds to the gravitational waveform in the Fourier domain without cosmic acceleration.
%
%
The above $\psiaccel$ matches with the ones found in Refs.~\cite{setoDECIGO, takahashinakamura}.
A term with $x^n$ represents the $n$-th post-Newtonian (PN) order relative to the leading Newtonian phase $\Psi_N (f)$, hence this is a ``-4PN'' correction.
For $\tilde{h}(f) \big|_{\mathrm{no \ accel}}$, we use the restricted-2PN waveform including spin-orbit coupling at 1.5PN order~\footnote{We do not take the spin-spin coupling at 2PN order into account since it has been shown that its contribution is negligible~\cite{cutlerharms}.} which is given in Eq.~(26)~\footnote{We set $\bar{\omega}$, $\beta_g$, $I_e$ and $\sigma$ to 0.} of Ref.~\cite{yagiLISA}, where ``restricted'' means that we only take the leading Newtonian quadrupole contribution to the amplitude and neglect contributions from higher harmonics.
%
%
%
%
%
%

\textit{Numerical Setups}:
The measurement accuracy on $\zdrift$ has been estimated mainly using ultimate DECIGO (which is three orders of magnitude more sensitive than DECIGO)~\cite{setoDECIGO,takahashinakamura}. 
In this letter, we improve their analyses in the following way and apply the result to probe the inhomogeneity of the universe:
(I) We use DECIGO/BBO, including the confusion noises from white dwarf (WD) binaries.
(II) Rather than sky-averaged analysis, we perform Monte Carlo simulations by randomly distributing the directions and orientations of sources. 
(III) We include the spin-orbit coupling into binary parameters.
(IV) We use the merger rate that reflects the star formation history rather than the redshift independent rate.
(V) We do not require that the source redshift $z$ is known by identifying the host galaxies with EM observations.  
 (VI) Since we are only interested in the positivity of the redshift drift, we consider the cumulative acceleration parameter $\cum (z)$ defined as 
\be
\cum  (z_k) \equiv \sum_{n=1}^k X_H(z_n)\,,
\label{cumX}
\ee
where $X_H(z) \equiv X(z)/H_0$ and $z_n = (n-0.5)\delta z$ with $\delta z$ representing the size of the redshift bin.

Here, we take the binary parameters as
\be
\theta^i = \left( \ln \mcz, \ln \eta, \beta, t_c, \phi_c, D_L, \bar{\theta}_{\mrm{S}}, \bar{\phi}_{\mrm{S}}, \bar{\theta}_{\mrm{L}}, \bar{\phi}_{\mrm{L}}, X_H \right).
\ee
$\beta$ is the spin-orbit coupling parameter while $\phi_c$ represents the coalescence phase and $D_L$ denotes the luminosity distance.
$(\bar{\theta}_{\mrm{S}}, \bar{\phi}_{\mrm{S}})$ give the direction of the source in the barycentric frame which is tied to the ecliptic and centered in the solar system barycenter (see Fig. 1 of Ref.~\cite{yagiLISA}), and $(\bar{\theta}_{\mrm{L}}, \bar{\phi}_{\mrm{L}})$ specify the orientation of the source orbital axis in the barycentric frame.

We estimate the measurement accuracies of binary parameters $\theta^i$ using Fisher analysis.
Assuming that the detector noise is stationary and Gaussian, the measurement accuracy is given as $ \Delta\theta^i  \equiv \sqrt{ \left( \Gamma^{-1}  \right)_{ii}}$~\cite{cutlerflanagan}.
%
%
Here, the Fisher matrix $\Gamma_{ij}$ is defined as $\Gamma_{ij}\equiv (\partial_ih|\partial_jh)$, where the inner product is defined as 
%
%
%
\be
(A|B) \equiv 4 \mathrm{Re}\int ^{f_{\mrm{fin}}}_{f_{\mrm{in}}}df \, \frac{\tilde{A}^{*}(f)\tilde{B}(f)}{S_n(f)}, 
\label{scalar-prod}
\ee
with $S_n(f)$ denoting the noise spectrum of a single interferometer and $f_{\mrm{in}}$ and $f_{\mrm{fin}}$ representing the initial and final frequencies of observation.
The measurement error of $\cum (z_k)$ can be estimated as 
\be
\Delta (\cum (z_k))  \equiv \left[ \sum_{n=1}^k \left( \Delta X_H(z_n) \right)^2 \right]^{1/2}.
\ee

We use the instrumental noise spectrum of BBO~\footnote{We assume that DECIGO also has the same sensitivity as BBO. The (non sky-averaged) instrumental noise spectral density for BBO with a single interferometer is given in Eq. (34) of Ref.~\cite{yagi:brane}. (Here, some typos have been corrected in the latest arXiv version.) The total noise spectral density including WD/WD confusion noises is shown in Eq. (36) of Ref.~\cite{yagi:brane}. In this letter, we assume that NS/NS foreground noise can be subtracted down to the level below the instrumental noise~\cite{cutlerharms}. } 
and also consider WD/WD confusion noises which deteriorate the sensitivity below $f\sim 0.2$Hz.
In this letter, we are interested in the phase shift due to the redshift drift, which appears at relative ``-4PN'' order.
The negative PN order indicates that this effect is larger when GW frequency is lower.
Since the WD/WD confusion noise would mask the lower frequency part of DECIGO/BBO observation window, this effect cannot be neglected.
We choose $f_{\mrm{in}}$ and $f_\mrm{fin}$ in Eq.~\eqref{scalar-prod} as $f_{\mrm{in}}=(256/5)^{-3/8} \pi^{-1} \mcz^{-5/8} \Delta t_o^{-3/8}$ and $f_{\mrm{fin}}=100$Hz, which correspond to the frequency at $\Delta t_o$ before coalescence and the higher cutoff frequency of the detector, respectively~\cite{yagi:brane}. 
For fiducial values, we set $m_1=m_2= 1.4M_{\odot}$ and take $t_c=\phi_c=\beta=0$.
We assume that a flat $\Lambda$CDM is the correct model and set the fiducial values as $H_0=70$km/s/Mpc and the cosmological parameters as $\Omega_m=0.3$ and $\Omega_{\Lambda}=0.7$.

The number of binaries $\Delta N(z)$  that exists in each redshift bin with the size $\delta z$ is estimated as $ \Delta N(z) =  4\pi \left[ a_0 r(z) \right]^2 \dot{n}(z) (d\tau /dz) \delta z \Delta t_o$~\cite{cutlerharms}, where $a_0r(z) =\int ^z_0 dz'/H(z')$ and ${d\tau}/{dz} = \{ (1+z)H(z) \}^{-1}$ with $H(z) \equiv H_0 \sqrt{\Omega_m (1+z)^3+\Omega_{\Lambda}}$. $a_0$, $r(z)$, and $\tau$ represent current scale factor, comoving distance and proper look back time, respectively. 
$\dot{n}(z)=\dot{n}_0 R(z)$ shows the NS/NS merger rate per unit comoving volume per unit proper time, where we assume the merger rate today as $\dot{n}_0=10^{-6}$ Mpc$^{-3}$ yr$^{-1}$~\cite{abadie} and 
%
\ba
R(z)=\left\{ \begin{array}{ll}
1+2z & (z\leq 1) \\
\frac{3}{4}(5-z) & (1\leq z\leq 5) \\
0 & (z\geq 5). \\
\end{array} \right.
\ea
The merger rate evolution against $z$ reflects the current observation of star formation history~\cite{schneider}.

For each fiducial redshift $z_n$ with the redshift bin size set as $\delta z =0.1$,
following Refs.~\cite{bertibuonanno,yagiLISA,yagi:brane}, we randomly generate $10^4$ sets of $(\bar{\theta}_{\mrm{S}},\bar{\phi}_{\mrm{S}},\bar{\theta}_{\mrm{L}},\bar{\phi}_{\mrm{L}})$ and for each set, we calculate $\sqrt{\left( \Gamma^{-1} \right)_{ii}}$. 
Then, we take the average to yield $\left[\sqrt{\left( \Gamma^{-1} \right)_{ii}}\right]_{\mrm{ave}}$. 
The measurement accuracy at each $z_n$ is estimated as
%
\be
\Delta\theta^i  = N_{\mrm{int}}^{-1/2} \Delta N(z_n)^{-1/2}  \left[\left( \Gamma^{-1} \right)^{1/2}_{ii}\right]_{\mrm{ave}},
\ee
where $N_{\mrm{int}}=8$ represents the effective number of interferometers~\cite{cutlerholz}.
DECIGO/BBO has four clusters of triangular detectors (see e.g. Fig. 2 of Ref.~\cite{yagi:brane} for the proposed configurations of DECIGO/BBO) and for simplicity, we assume that all the clusters are placed on the same site.
First, we show a rough estimate of the measurement accuracy of $X_H$ using DECIGO/BBO and then show our numerical results.

\textit{Rough Estimate}:
Let us assume that we observe $\Delta N (z=1) = 2.4 \times 10^5$ NS/NS binaries for 5yr observations. 
The measurement accuracy of $X_H(z)$ is mostly determined at the lowest frequency of the observation, i.e. at $f=f_{5\mrm{yr}}=0.073$Hz since the effect due to the cosmic acceleration is larger on GW signals with lower frequency.
We first define the squared signal-to-noise ratio (SNR) as $\mrm{SNR}^2 \equiv \int_{f_\mrm{in}}^{f_\mrm{fin}} \rho^2(f) d\ln f $, where $\rho(f)^2 \equiv 4 N_\mrm{int} f |\tilde{h}(f)|^2/S_n(f) $ 
expresses the contribution to SNR$^2$ at each frequency $f$.
Typical values for $\rho(f_{\mathrm{in}})$ and $X_H$ at this redshift would be $\rho(f_{\mathrm{in}}) \approx 1.6$ and $X_H(z=1) \approx 0.05$, respectively.
Roughly speaking, 
if $\psiaccel \rho(f_{\mathrm{in}}) \sqrt{\Delta N} $ exceeds $\mathcal{O}(1)$,
the effect due to the cosmic acceleration can be detected (when we assume that there is no correlation between $\zdrift$ and other parameters).
Therefore, we have the measurement accuracy of $ \Delta X_H (z=1)  \approx 0.05$ which is comparable to the fiducial value of $X_H(z=1)$. 
Hence, it is marginal whether the positivity of the redshift drift can be measured or not.
We expect that the measurement accuracy would be improved when we consider $\cum (z_k)$.

\begin{figure}[t]
  \centerline{\includegraphics[scale=1.4,clip]{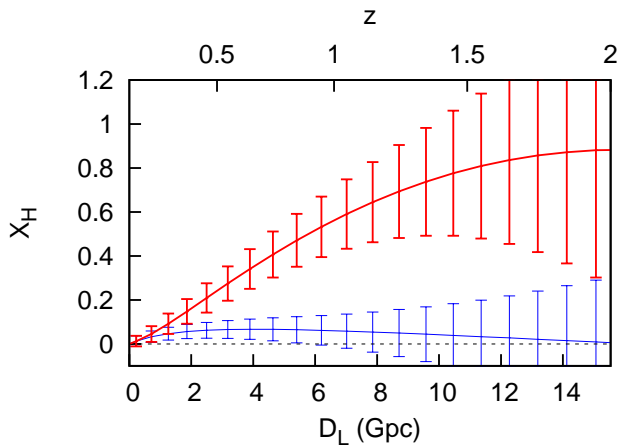} }
 \caption{\label{drift-5yr}  
The acceleration parameter $X_H$ (blue thin), the cumulative acceleration parameter $\cum $ (red thick) and their measurement accuracies under $\Lambda$CDM model with 5yr observations using DECIGO/BBO. 
On the horizontal axes, we show both $D_L$ and $z$.
The former can be measured from GW observations while the latter is degenerate with mass parameters.}
\end{figure}

\textit{Numerical Results}:
In Fig.~\ref{drift-5yr}, the (blue) thin line shows $X_H$ and its measurement accuracy for 5yr observations with DECIGO/BBO.
In GW observations, since the redshift $z$ is degenerate with mass parameters~\footnote{These degeneracies can be solved via tidal effect if we know the equations of state of a NS \textit{a priori}~\cite{read}. They can also be disentangled by using the observed chirp mass distribution of NSs~\cite{taylor}. }, we show the result against $D_L$, with corresponding $z$ just for reference.
We emphasize that the test can be performed with GW observations alone by using the $D_L$--$X_H$ relation.
It can be seen that at $z \approx 0.5$,  DECIGO and BBO can marginally probe the positivity of the redshift drift.
Our result is consistent with the rough estimate explained in the previous paragraph.
This test can be improved, shown by the (red) thick lines, by considering the cumulative acceleration parameter $\cum $.
By using this quantity, we can probe the positivity of the redshift drift with sufficient accuracy.

\begin{figure}[t]
  \centerline{\includegraphics[scale=1.4,clip]{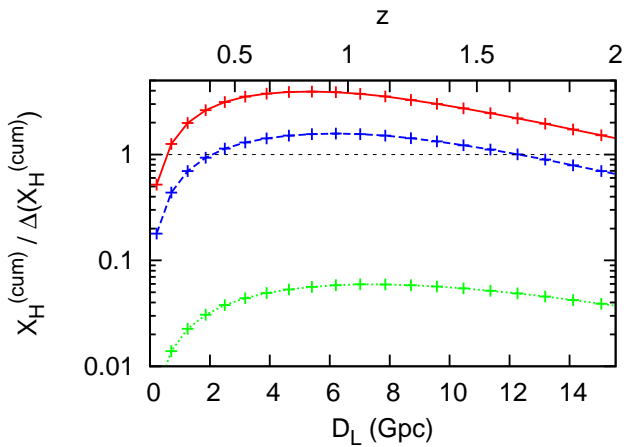} }
 \caption{\label{drift-ratio}
The ratio between the fiducial values of $\cum $ and its measurement accuracies for 1yr (green dotted), 3yr (blue dashed) and 5yr (red solid) observations. 
This shows to what $\sigma$-level we can detect the positivity of the redshift drift with.
 }
\end{figure}

Figure~\ref{drift-ratio} shows the ratio between the fiducial value of $\cum $ and its 1-$\sigma$ measurement error.
We can see that it is possible to measure the positivity of $\zdrift$ with 
1.5-$\sigma$ for 3yr observations and 4-$\sigma$ for 5yr observations.
Furthermore, with 5yr observations, it may even be possible to probe the positivity of $\zdrift$ at around $z\approx 0.2$ with 2-$\sigma$ confidence.

\textit{Conclusions and Discussions}:
In this letter, we have estimated how accurately 
we can directly measure the acceleration of the universe 
with future space-borne GW interferometers such as DECIGO and BBO.
Assuming that the $\Lambda$CDM model is correct,
we have shown that, with 3--5yr observations, 
we will be able to measure the positivity of the redshift drift, 
which enables us to rule out \textit{any} LTB void model 
with a monotonically increasing density profile.
Also,  with 5yr observations, 
DECIGO/BBO can measure the positivity of the redshift drift at even $z\sim 0.2$.
This indicates that future GW observations will be able to 
rule out \textit{any} LTB void model unless 
we allow unrealistically steep density gradient at $z \sim$0. 
Our results regarding the use of DECIGO/BBO are complementary 
to the ones 
proposed by Quartin and Amendola~\cite{quartin} where 
EM observations are used in the sense that these observations target different sources at different redshift ranges.
However, GW observations can be more powerful since they can offer a more generic test of the LTB model.

A variety of dark energy models predict similar values of $\zdrift$ 
to the $\Lambda$CDM model (see Quartin and Amendola~\cite{quartin} and references therein).
Therefore, we expect that the test can be performed with almost 
the same accuracy even if the fiducial model is another dark energy model.

Unfortunately, it seems very difficult to measure $\zdrift$ 
with ground-based detectors, even with third-generation ones like Einstein Telescope (ET)~\cite{et}.
The advantages of using DECIGO/BBO over ground-based ones are 
(i) they have a larger number of GW cycles, 
(ii) the effect of $\zdrift$ is ``-4PN'', meaning that we can perform our test better 
with lower-frequency GWs,  
(iii) they have a longer observation time for each binary, and 
(iv) they detect a larger number of NS binaries. 

We need to comment about the peculiar acceleration of each binary source. 
This 
acts as an additional ``noise'' 
when measuring the redshift drift. 
Amendola \etal~\cite{amendola:peculiar} have estimated the peculiar 
accelerations for typical clusters and galaxies, 
and found that they are almost the same 
magnitude as the cosmological acceleration. 
However, they are much smaller than the measurement errors of 
$\zdrift$ from a single binary source due to detector noises, hence we can safely neglect the effect of peculiar accelerations (see also Uzan \etal~\cite{uzan:peculiar}).

In this letter, we have assumed that the 
binary orbits
are circular. 
If we include the eccentricity, 
this may be degenerate with $\zdrift$ 
since both have 
large effects 
when the binary separation is large. 
Thus, in future work, we need to estimate how accurately we can determine 
the positivity of $\zdrift$ in eccentric binaries.
Also, we have used $X_H^{\mrm{(cum)}}$ as our estimator because of its simple form, but it may not be the best one.
It would be interesting to find the corresponding one that performs better in probing the positivity of the redshift drift.
Furthermore, we note here that we have used 
the LTB metric as a simple effective model. 
To be more realistic, we need to use more sophisticated models such as the Swiss-Cheese model~\cite{swisscheese} (see a recent discussion by C\'el\'erier~\cite{celerier}). 
We leave these issues for future work.

We thank Takahiro Tanaka, Naoki Seto and Takashi Nakamura for discussions and valuable comments.
We also thank Daniel Holz and Jonathan White for carefully reading this manuscript and giving us useful advice.
KY, AN and CY are supported by a Grant-in-Aid through the Japan Society for the Promotion of Science (JSPS).

\bibliography{ref}

\end{document}